\documentclass{anstrans}
\title{A Multi-Criteria Evaluation Framework for Siting Fusion Energy Facilities: Application and Evaluation of U.S. Coal Power Plants

}
\author{Muhammad R. Abdussami,$^{*,{\dagger}}$ Kevin Daley,$^{*,{\dagger}}$, Gabrielle Hoelzle,$^{*,{\dagger}}$ and Aditi Verma,$^{,{\dagger}*}$}

\institute{
$^{*}$Fastest Path to Zero Initiative, University of Michigan, Ann Arbor, MI, United States,

$^{\dagger}$Department of Nuclear Engineering and Radiological Sciences, University of Michigan, 2355 Bonisteel Blvd, Ann Arbor, MI 48109, United States, rafiul@umich.edu, aditive@umich.edu.

}

\usepackage{graphicx} 
\usepackage{booktabs} 
\usepackage{microtype} 

\usepackage{url}

\begin{document}
\section{Introduction}

As fusion energy approaches commercial readiness, siting demonstration and future power plant facilities have become a priority for technology developers. Several private and public-sector initiatives, such as Zap Energy \cite{FusionZAPCoal}, UK government \cite{FusionUKCoal}, Type One Energy Group \cite{FusionTVACoal}, and Commonwealth Fusion Systems (CFS) \cite{FusionCFSCoal}, have already begun identifying potential sites for first-of-a-kind fusion plants, with a notable trend toward reusing retired coal plant infrastructure. These early projects suggest that former coal sites may offer strategic advantages—such as grid connectivity, land availability, and a potentially interested and supportive host community—all of which are conducive to building first-of-a-kind facilities on projected timelines and budgets.

While coal-to-fission transitions have been extensively studied, the use of fusion technology for repurposing coal sites remains underexplored. Fusion offers some distinct advantages over fission, including inherent safety, negligible long-lived radioactive waste, and widely available fuel sources, all of which may lead to a more positive public sentiment toward fusion facilities and fewer regulatory barriers. Pilot efforts—such as Zap Energy’s Z-pinch project at Centralia, Washington \cite{FusionZAPCoal}, and the UK STEP program at the West Burton coal site \cite{FusionUKCoal}—demonstrate the relevance of coal sites as candidate locations for fusion deployment and highlight the need for robust siting frameworks tailored to fusion technologies.

Repurposing coal plants with fission nuclear reactors has been demonstrated to be technically and economically viable in various national contexts \cite{abdussami2025future}. One study finds that up to 80\% of U.S. coal sites could support advanced reactor deployment when considering siting, cost, and economic factors \cite{hansen2022investigating}. In South Korea, evaluations demonstrate that Small Modular Reactors (SMRs) offer compatibility with existing coal infrastructure and provide grid stability advantages over renewables \cite{joo2024evaluation}.

This paper proposes a comprehensive methodology for siting fusion energy facilities, integrating expert judgment, geospatial data, and multi-criteria decision-making tools to evaluate site suitability systematically. As a case study, we apply this framework to all currently operational coal power plant sites in the United States to examine their potential for hosting future fusion facilities at a time when these coal plants are shut down on reaching their end of life -- timelines which are expected to coincide with the potential deployment of fusion energy facilities. Drawing on 22 siting criteria—including state and federal policies, risk and hazard assessments, and spatial and infrastructural parameters—we implement two Multi-Criteria Decision-Making (MCDM) methods: the Fuzzy Full Consistency Method (F-FUCOM) to derive attribute weights and the Weighted Sum Method (WSM) to rank sites based on composite suitability scores. By focusing on fusion-specific siting needs and demonstrating the framework through a coal site application, this study contributes a scalable and transparent decision-support tool for identifying optimal fusion energy deployment locations.

\section{Research Methodology}

\begin{figure*}[htbp]
\centering
\includegraphics[width=\textwidth, height=0.9\textheight, keepaspectratio]{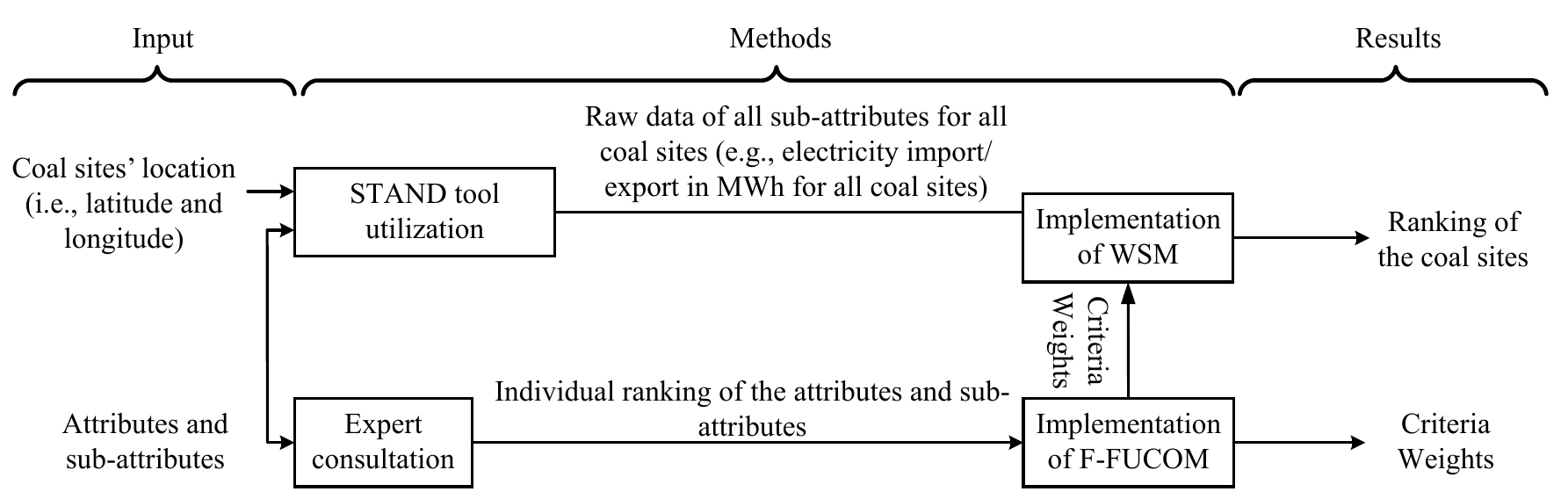}
\caption{Flowchart of research methodology.}\label{fig: Flowchart of research methodology}
\end{figure*}

\begin{figure*}[htbp]
\centering
\includegraphics[width=\textwidth, height=0.9\textheight, keepaspectratio]{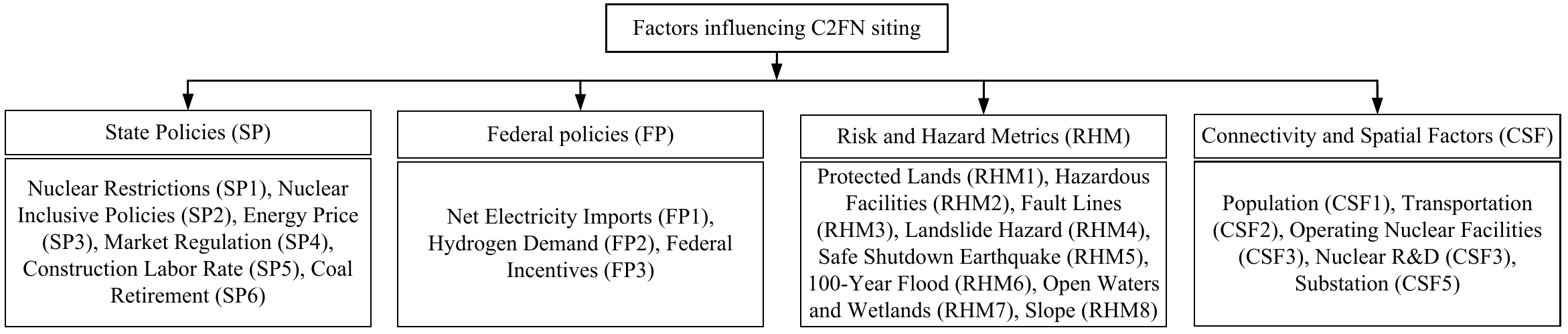}
\caption{List of attributes and sub-attributes for coal-to-fusion siting.}\label{fig: list of attribute}
\end{figure*}

The research methodology, illustrated in Fig.~\ref{fig: Flowchart of research methodology}, employs a structured, multi-criteria approach to evaluate the suitability of 220 U.S. operational coal plant sites (excluding those in Alaska and Puerto Rico due to data unavailability) for fusion energy deployment. First, the locations of all candidate coal sites were identified, and relevant sub-attribute data, such as energy prices, electricity import/export, etc., were listed. Then, the Siting Tool for Advanced Nuclear Development (STAND) was used to process coal site locations and associated sub-attributes to generate raw input data for this analysis \cite{abdussami2024investigation}. Binary values (e.g., the presence of open water or wetlands) were converted to a numerical form for computation, with True assigned the value 1 and False assigned the value 0. We developed four thematic categories of attributes: State Policies (SP), Federal Policies (FP), Risk \& Hazard Metric (RHM), and Connectivity \& Spatial Factors (CSF), with each of the 22 sub-attributes being assigned to one of the four attribute categories in Fig.~\ref{fig: list of attribute}

At the outset of this study, five fusion experts from governmental and non-governmental organizations completed a survey in which they rated the relative importance of each sub-attribute on a scale from 1 (Not Important) to 5 (Extremely Important). These survey responses were then discussed in follow-up interviews to further explore the reasoning behind their assessments. The F-FUCOM method was then applied. F-FUCOM was used to assign consistent weights to criteria by incorporating expert judgment through fuzzy logic, which handles uncertainty and reduces the number of required comparisons compared to traditional methods \cite{ilieva2020fuzzy}. The F-FUCOM derives consistent weights for the decision criteria based on expert rankings and fuzzy linguistic comparisons. These inputs are translated into fuzzy numbers, and an optimization model, shown in Eq.~\ref{eq: FUCOM10}, is solved for each fusion expert to ensure consistency, $\chi$, and transitivity in the weighting process \cite{pamucar2020prioritizing}. The final output is a set of optimized fuzzy weights that reliably capture expert judgment while maintaining full logical consistency. The weights of the attributes are determined using FUCOM rather than F-FUCOM because of the nature of the experts' input.

\begin{equation}
\begin{array}{c}
\text{minimize} \quad \chi \\[1ex]
\text{s.t.} \quad 
\left\{
\begin{array}{l}
\left| w_k - w_{k+1} \otimes \varphi_{k/(k+1)} \right| \leq \chi, \quad \forall j \\[1ex]
\left| w_k - w_{k+1} \otimes \varphi_{k/(k+1)} \otimes \varphi_{(k+1)/(k+2)} \right| \leq \chi, \quad \forall j \\[1ex]
\sum\limits_{j=1}^{n} w_j = 1, \quad \forall j \\[1ex]
w_j^l \leq w_j^m \leq w_j^u \\[1ex]
w_j^l \geq 0, \quad \forall j \\[1ex]
j = 1, 2, \ldots, n
\end{array}
\right.
\end{array}
\label{eq: FUCOM10}
\end{equation}

where $w_k$ is the weight of the k-th criterion, $\varphi_{k/(k+1)}$ is the fuzzy comparative significance, ${w}_j = (w_j^l, w_j^m, w_j^u)$, and 
$\varphi_{k/(k+1)} = (\varphi_{k/(k+1)}^l,\ \varphi_{k/(k+1)}^m,\ \varphi_{k/(k+1)}^u)$.

These weights are then integrated with the site-level data using the WSM, which enables the final ranking of coal sites based on their overall suitability for repurposing as fusion energy sites. The suitability score of coal sites using WSM is determined by Eq.~\ref{eq:wsm}. 

\begin{equation}
A_i = X_{norm} W'
\label{eq:wsm}
\end{equation}
where $X_{norm}$ and $W'$ are the normalized decision matrix and weigh matrix of each sub-criteria, respectively.

\section{Results}

\subsection{Determination of criteria weights}

To determine the relative importance of key siting criteria, including State Policies (SP), Federal Policies (FP), Risk and Hazard Metrics (RHM), and Connectivity and Spatial Factors (CSF), expert input from five fusion energy experts is utilized. Each expert ranked the attributes based on their perceived importance, and the FUCOM method was employed to derive consistent priority weights. The derived weights achieve full consistency ($\chi$ = 0), with FP and CSF typically ranking high in priority. Optimization problems were formulated using expert-assigned pairwise comparative significance coefficients and solved using the Gurobi solver.

For sub-attribute analysis, a fuzzy version of FUCOM (F-FUCOM) was applied to six sub-criteria under the SP category. Experts provide linguistic assessments (e.g., “Extremely Significant” or “Moderately Significant”), which are converted into Triangular Fuzzy Numbers. After solving the F-FUCOM models, fuzzy and crisp weights are derived using the Graded Mean Integration Representation (GMIR) method. The results show that SP3 (Energy Price), SP6 (Coal Retirement), and SP2 (Nuclear Inclusive Policies) consistently receive higher weights, indicating their critical role in assessing the suitability of coal sites for fusion energy facility development.

For the FP category, expert inputs were collected to evaluate three sub-attributes: Net Electricity Imports (FP1), Hydrogen Demand (FP2), and Federal Incentives (FP3). The results indicate that FP3 is consistently ranked as the most influential factor by four out of five experts, followed by FP1. Full consistency ($\chi$ = 0) is achieved in all optimization problems. 

In the RHM category, eight sub-attributes are assessed. The results suggest a relatively balanced importance across many of the RHM sub-criteria, although fault lines (RHM3), landslide hazards (RHM4), and 100-year flood (RHM6)  are more frequently ranked as the most important. Besides, consistency levels are well within acceptable bounds ($\chi$ < 0.10).

For the CSF category, expert evaluations are collected to assess the importance of five sub-attributes: Population (CSF1), Transportation (CSF2), Operating Nuclear Facilities (CSF3), Nuclear R\&D (CSF4), and Substation (CSF5). The optimization results demonstrate a strong consensus among experts on the importance of transportation infrastructure (CSF2) and substation availability (CSF5). All consistency values ($\chi$) remain below 0.10, confirming that the models achieved an acceptable level of consistency.

\begin{figure}[htbp]
\centering
\includegraphics[width=0.5\textwidth, height=0.99\textheight, keepaspectratio]{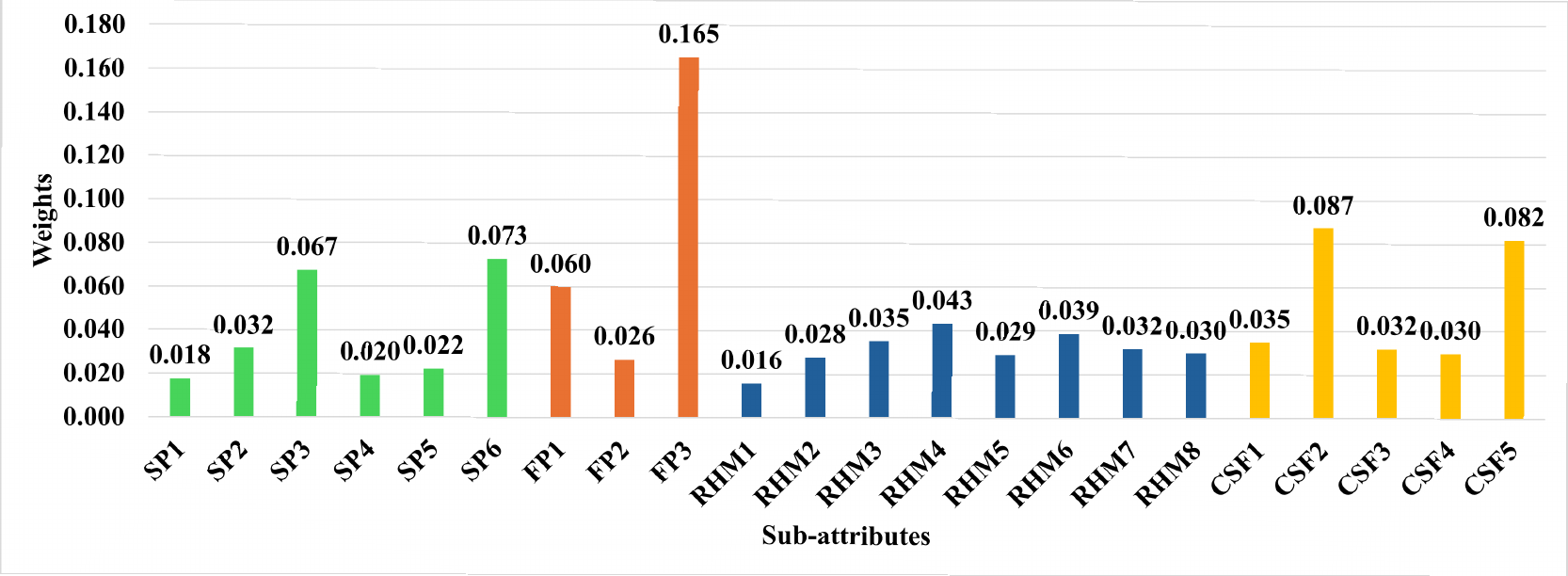}
\caption{Weights of the sub-attribute.}\label{fig: Weights of the sub-attribute}
\end{figure}

The global weights of all sub-attributes are determined by multiplying each sub-criterion’s crisp weight with its corresponding criteria weight and averaging across expert inputs. This approach enables the identification of both the most and least influential factors in the coal-to-fusion site selection process.
Among the four main criteria, CSF emerges as the most influential, with a total weight of 26.59\%, followed closely by FP at 25.16\%, RHM at 25.09\%, and SP at 23.15\%, as shown in Fig.~\ref{fig: Weights of the sub-attribute}. This indicates a relatively balanced contribution from all major siting domains.
At the sub-attribute level, the top five most influential factors are identified as: FP3 (Federal Incentives, 16.52\%), CSF2 (Transportation, 8.73\%), CSF5 (Substation, 8.20\%), SP6 (Coal Retirement, 7.29\%), and SP3 (Energy Price, 6.74\%).

Conversely, RHM1 (Protected Lands), SP1 (Nuclear Restrictions), and SP4 (Market Regulation) are rated as the least influential, with weights of 1.57\%, 1.76\%, and 1.96\%, respectively.

\subsection{Determination of the ranking of the coal sites}

\begin{figure}[htbp]
\centering
\includegraphics[width=0.4\textwidth, height=0.9\textheight, keepaspectratio]{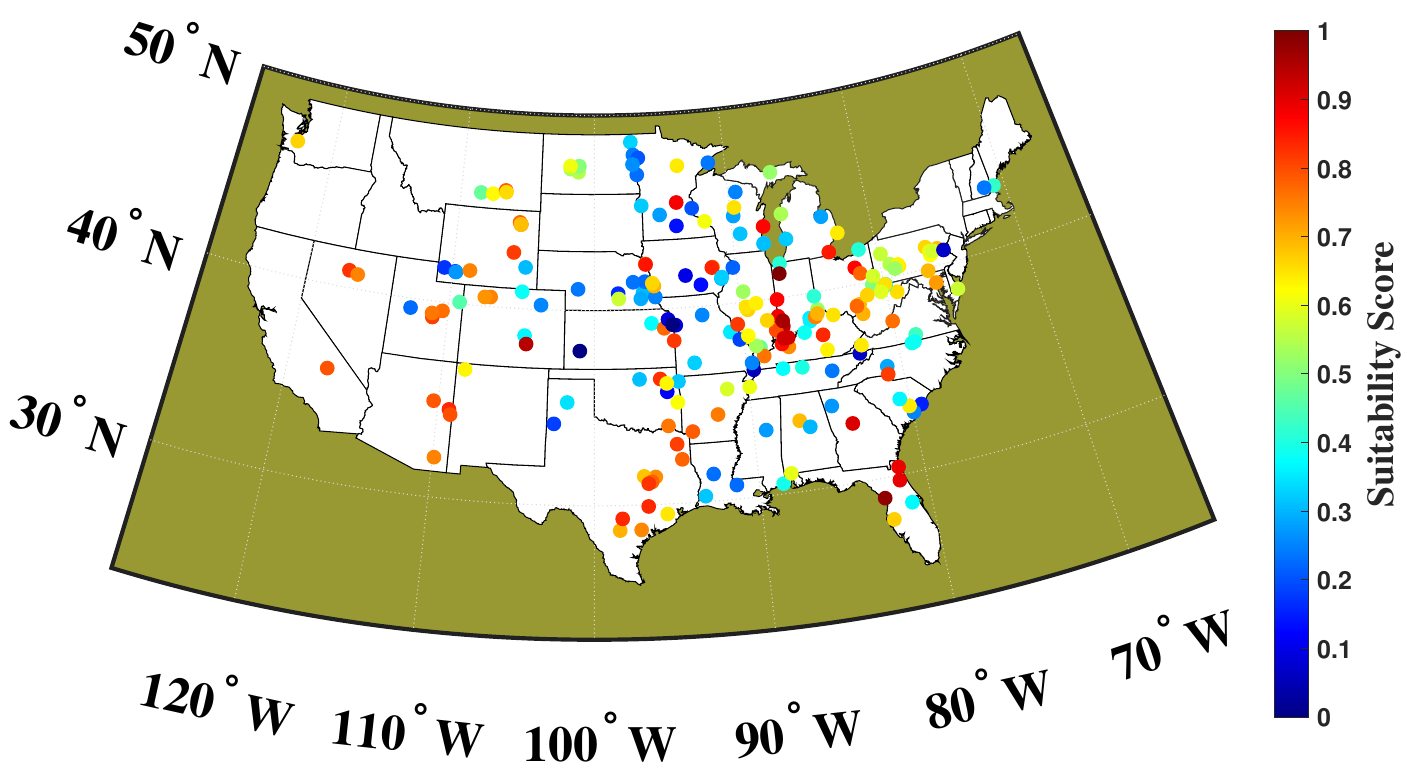}
\caption{All the studied coal sites and their corresponding suitability scores (normalized).}\label{fig: All the studied coal sites and their corresponding suitability scores (normalized)}
\end{figure}

Using the previously determined sub-criteria weights, the WSM method was applied to rank 220 coal plant sites across the United States based on their suitability for fusion energy deployment. The suitability score for each coal site, as determined by WSM, is presented in Fig.~\ref{fig: All the studied coal sites and their corresponding suitability scores (normalized)}. The R M Schahfer plant in Indiana emerges as the top-ranked site overall, while the Holcomb in Kansas ranks as the lowest suitable site.

\begin{figure}[htbp]
\centering
\includegraphics[width=0.4\textwidth, height=0.9\textheight, keepaspectratio]{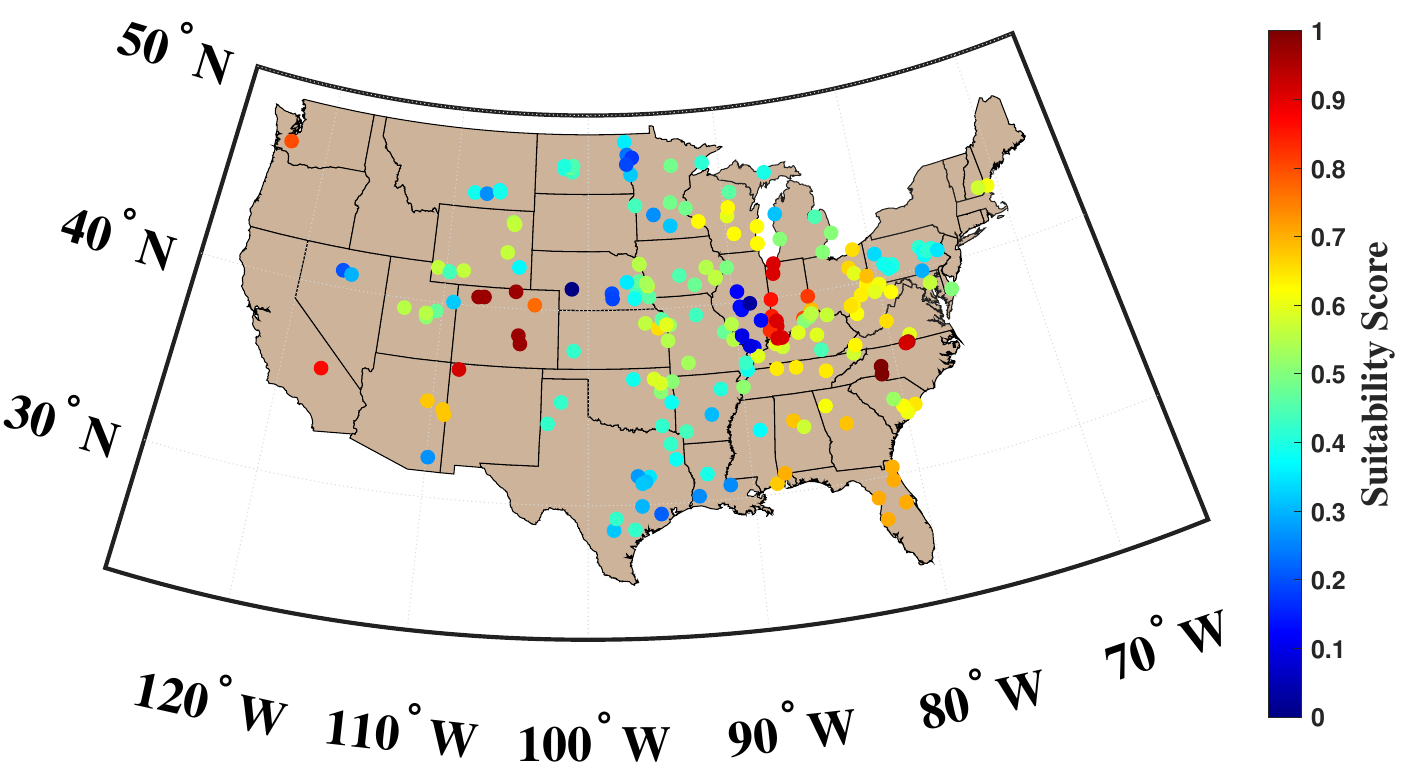}
\caption{Suitability score of the coal sites based on only SP (normalized).}\label{fig: Suitability score of the coal sites based on only SP}
\end{figure}

\begin{figure}[htbp]
\centering
\includegraphics[width=0.4\textwidth, height=0.5\textheight, keepaspectratio]{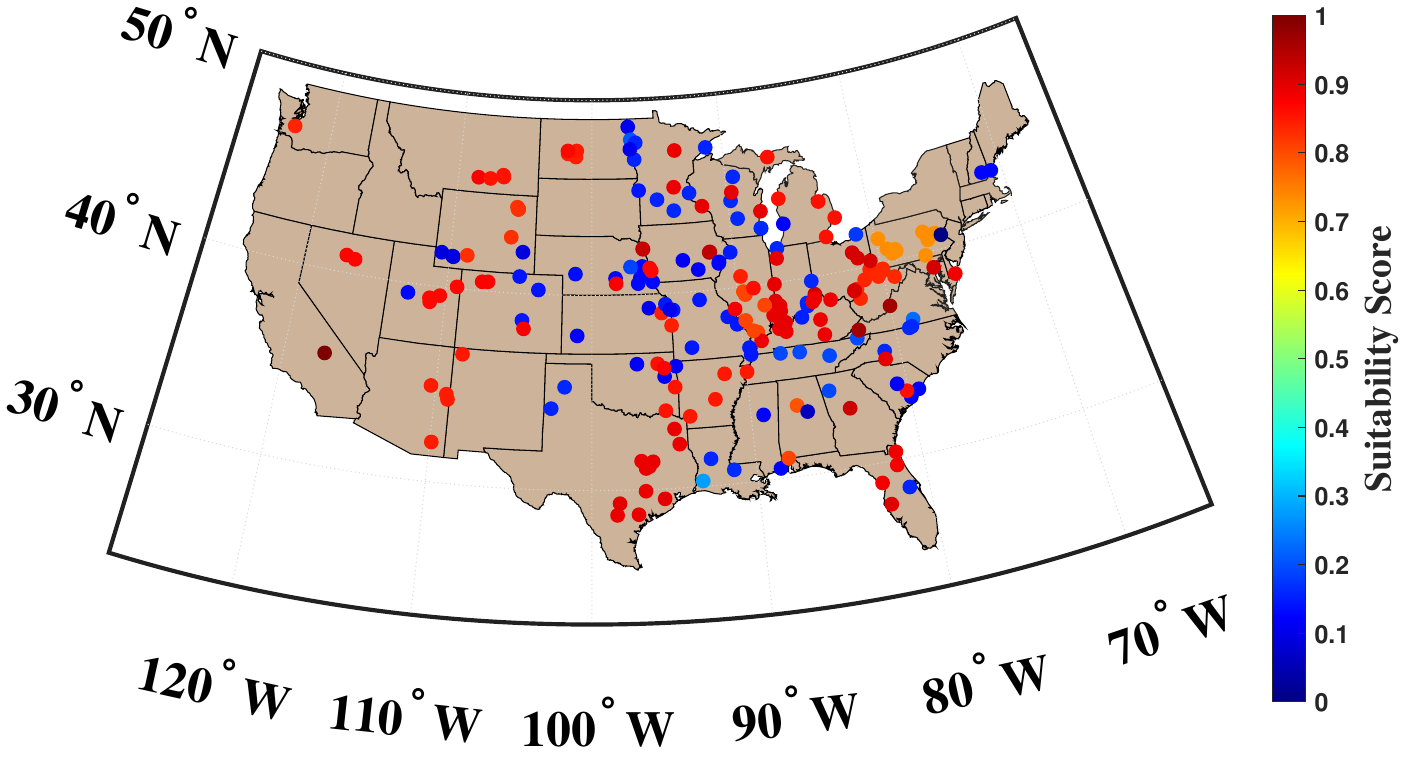}
\caption{Suitability score of the coal sites based on only FP (normalized).}\label{fig: Suitability score of the coal sites based on only FP}
\end{figure}

\begin{figure}[htbp]
\centering
\includegraphics[width=0.4\textwidth, height=0.5\textheight, keepaspectratio]{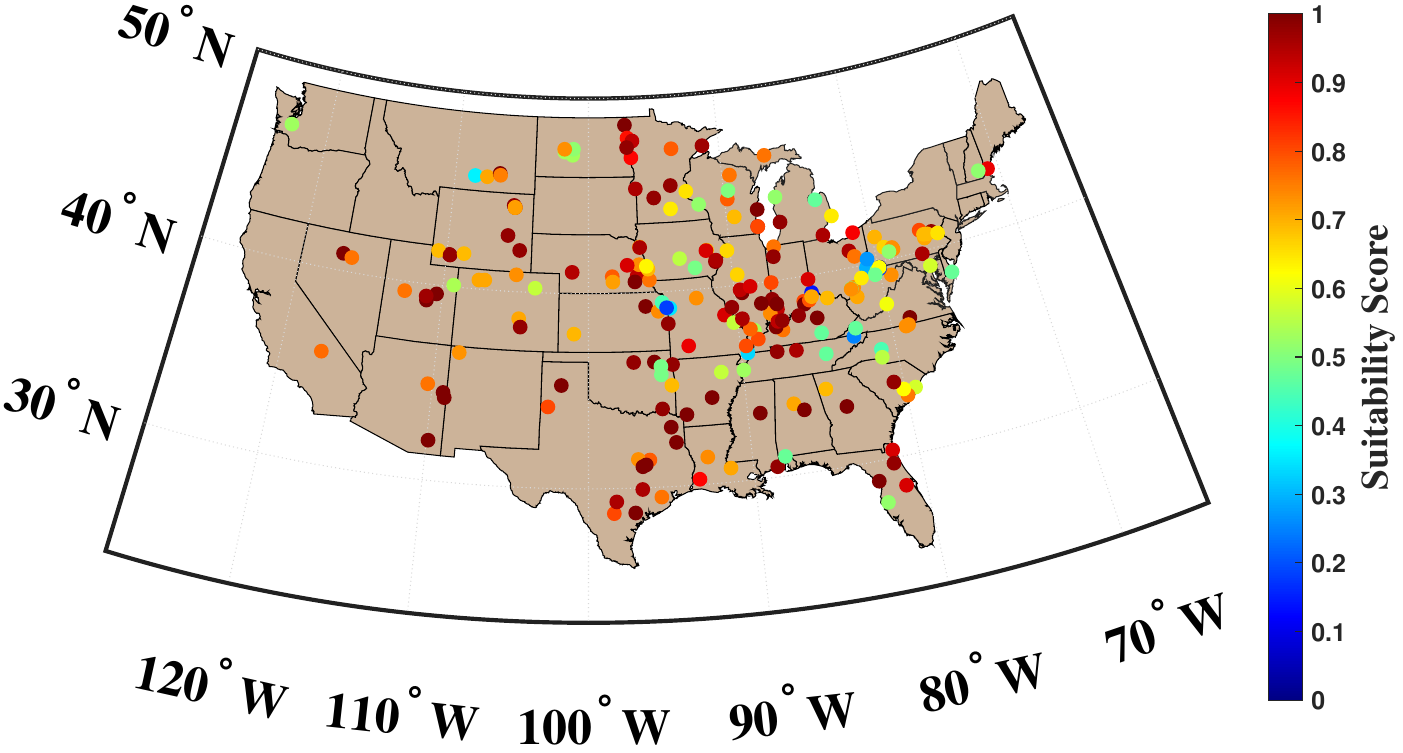}
\caption{Suitability score of the coal sites based on only RHM (normalized).}\label{fig: Suitability score of the coal sites based on only RHM}
\end{figure}

\begin{figure}[htbp]
\centering
\includegraphics[width=0.44\textwidth, height=0.5\textheight, keepaspectratio]{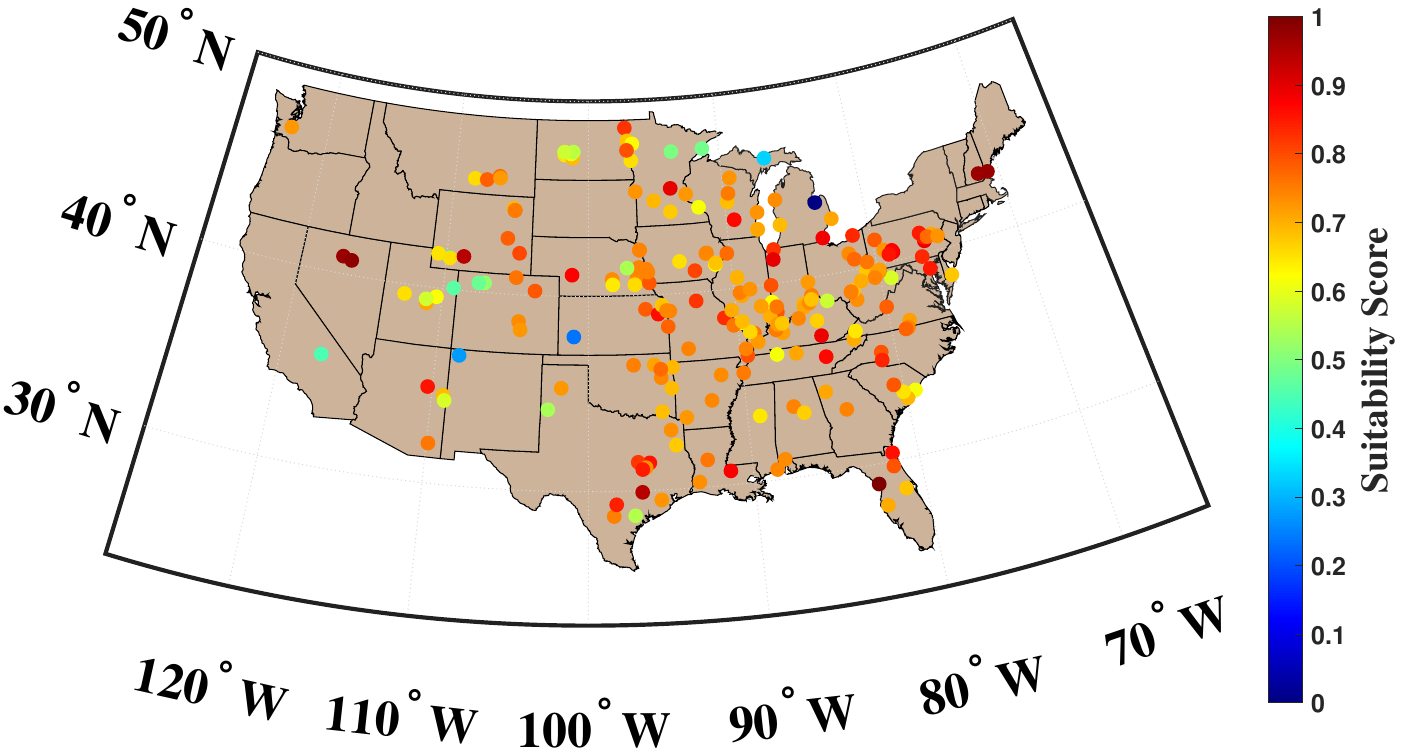}
\caption{Suitability score of the coal sites based on only CSF (normalized).}\label{fig: Suitability score of the coal sites based on only CSF}
\end{figure}

To better understand site performance, additional analyses are conducted for each major attribute group: SP, FP, RHM, and CSF, presented in Fig.~\ref{fig: Suitability score of the coal sites based on only SP}, Fig.~\ref{fig: Suitability score of the coal sites based on only FP}, Fig.~\ref{fig: Suitability score of the coal sites based on only RHM}, and Fig.~\ref{fig: Suitability score of the coal sites based on only CSF}, respectively. Results show that some top-ranked sites in overall suitability may not rank highest when evaluated under individual attributes. For instance, while R M Schahfer ranks first overall, its sub-ranks are 13th (SP), 25th (FP), 36th (RHM), and 9th (CSF), suggesting attribute-specific trade-offs.
Other top-performing sites by attribute include: Marshall (NC) in SP, Argus Cogen (CA) in FP, Welsh (TX) in RHM, and Crystal River (FL) in CSF.

These results emphasize the importance of a holistic, multi-attribute approach to understanding the feasibility of the coal-to-fusion study. They also provide actionable insights for policymakers, indicating that different sites may be better suited depending on which criteria are prioritized for siting fusion energy facilities.

\section{Conclusions}

This study presents a comprehensive and data-driven framework for evaluating the suitability of retired coal plant sites across the United States for the deployment of fusion energy. By integrating expert judgment with structured multi-criteria decision-making methods, including F-FUCOM and WSM, we assess the relative importance of technical, policy, environmental, and infrastructural factors influencing the potential for siting fusion energy facilities at retiring coal plant sites. Results indicate that while all four main criteria — SP, FP, RHM, and CSF — contribute significantly to site suitability, CSF plays a particularly dominant role. Sub-attributes such as federal incentives, transportation, substations, coal retirement timelines, and energy prices emerge as key drivers in the siting process.
The application of the WSM methodology to 220 coal sites reveals meaningful spatial patterns. It highlights specific facilities, such as Marshall (NC), Argus Cogen (CA), Welsh (TX), and Crystal River (FL), as top candidates under SP, FP, RHM, and CSF evaluation, respectively. The divergence between overall and attribute-specific rankings further reinforces the need for more granular-level assessment before the practical deployment of fusion reactors. It is important to emphasize that the site assessment methodology described here alone is insufficient for site selection. Community and stakeholder engagement, particularly with host communities and at the state level, will provide socially and locally grounded insights and perspectives that are necessary for making fair and informed siting decisions. 

Future work will extend this framework by incorporating dynamic parameters such as evolving energy market conditions, community sentiment toward fusion, community sentiment toward coal, community sentiment toward energy \cite{diem2025socially}, and climate resilience factors. Besides, a comprehensive sensitivity analysis of the sub-attributes will be conducted to determine the impact of each sub-attribute on the likelihood of rank reversal. Finally, integrating lifecycle cost assessments and techno-economic modeling of fusion technologies will further refine prioritization for demonstration and commercial deployment phases.

\section{Acknowledgments}
This work is sponsored by the Department of Energy Office of Nuclear Energy under project number (DE-NE0009382), which is funded through the Nuclear Energy University Program (NEUP). We are also grateful to the five fusion experts for their valuable insights and participation in the survey and interview process.

\bibliographystyle{ans}
\bibliography{bibliography}
\end{document}